# Evaluation of various Deformable Image Registrations for Point and Volume Variations

Su Chul Han[1,2], Sang Hyun Choi[2], Seungwoo Park[2], Soon Sung Lee[1,2], Haijo Jung[2], Mi-Sook Kim[2], Hyung Jun Yoo[2], Young Hoon Ji[1,2] and Kum Bae Kim[1,2,*]

[1]*Department of Radiological Cancer Medicine, University of Science and Technology, Daejeon 305-350, Korea*

[2]*Research Center for Radiotherapy, Korea Institute of Radiological and Medical Sciences, Seoul 139-706, Korea*

Chul Young Yi[3]

[3]*Department of Ionizing Radiation Standards, Korea Research Institute of Standards and Science, Daejeon 305-340, Korea*

The accuracy of deformable image registration (DIR) has a significant dosimetric impact in radiation treatment planning. There have been many groups that have studied about accuracy of DIR. In this study, we evaluated accuracy of various DIR algorithms using variations of the deformation point and volume.

The reference image ($I_{ref}$) and volume ($V_{ref}$) was first generated with virtual deformation QA software (ImSimQA, Oncology System Limited, UK). We deformed $I_{ref}$ with axial movement of deformation point and $V_{ref}$ depending on the type of deformation (relaxation and contraction) in ImSimQA software. The deformed image ($I_{def}$) and volume ($V_{def}$) acquired by ImSimQA software were inversely deformed to $I_{ref}$ and $V_{ref}$ using DIR algorithms. As a result, we acquired deformed image ($I_{id}$) from $I_{def}$ and volume ($V_{id}$) from $V_{def}$. Four intensity-based algorithms were tested following that the horn–schunk optical flow (HS), iterative optical flow (IOF), modified demons (MD) and fast demons (FD) with the Deformable Image

Registration and Adaptive Radiotherapy Toolkit (DIRART) of MATLAB. The image similarity between $I_{ref}$ and $I_{id}$ was calculated to evaluate accuracy of DIR algorithms using the metrics that were Normalized Mutual Information (NMI) and Normalized Cross Correlation (NCC).

When moving distance of deformation point was 4 mm, the value of NMI was above 1.81 and NCC was above 0.99 in all DIR algorithms. Since the degree of deformation was increased, the degree of image similarity was decreased. When the $V_{ref}$ increased or decreased about 12%, the difference between $V_{ref}$ and $V_{id}$ was within ±5% regardless of the type of deformation which was classified into two types that are the deformation1 is to increase the $V_{ref}$ (relaxation) and the deformation 2 is to decrease the $V_{ref}$ (contraction). The value of Dice Similarity Coefficient (DSC) was above 0.95 in deformation1 except for the MD algorithm. In case of deformation 2, that of DSC was above 0.95 in all DIR algorithms. The $I_{def}$ and $V_{def}$ have not been completely restored to $I_{ref}$ and $V_{ref}$ and the accuracy of DIR algorithms was different depending on the degree of deformation. Hence, the performance of DIR algorithms should be verified for the desired applications

Email: tommikim77@gmail.com

Fax: +82-2-970-2412

## I. INTRODUCTION

The multi-fraction treatment of radiotherapy has the potential for deformation of tumors and normal tissues. Because of this, the miscalculation of cumulative doses for multi fraction treatments may be occurred. To solve this problem, Adaptive Radiotherapy (ART) has been studied [1]. Recently, Deformable Image Registration (DIR) has been a very considerable part in ART [2].

There are several commercially or publicly available DIR algorithms that have been applied to various medical applications. For example, Kessler used DIR algorithms for image fusion in multimodality [3] and Ragan et al. applied a deformable model to the semi-automated segmentation in Four-Dimensional Computed Tomography (4DCT) [4]. Zhang et al used DIR algorithms in lung functional (ventilation) imaging in thoracic cancer patients [5].

As the application of the DIR increases, it is highly required to provide an evaluation of the accuracy of their deformable image registration for the desired application. Wognum et al noted that small errors in the deformation map can result in significant changes in the bladder dose in areas with high dose gradients and it is necessary for high spatial accuracy of the DIR [6]. Kriby et al described that the accuracy of the DIR may have a significant dosimetric impact on radiation treatment planning. Thus, the quality assurance of DIR algorithms could be required during the treatment process [7].

To evaluate accuracy of DIR algorithms, there are several metrics using estimates of image similarity. For instance, the metrics that used only intensity were Sum of Squared intensity (SSD) and Sum of Absolute intensity difference (SAD), and Normalized Cross Correlation (NCC). Other metrics that based image information were joint entropy and Mutual Information (MI) [8]. The Dice Similarity Coefficient (DSC) and Tanimoto Coefficient (TC) were used to evaluate the degree of overlap between volumes [9]. For instance, Wognum et al used various metrics such as Surface Distance Error (SDE), Hausdorff Distance (HD) and

DSC for validation of DIR algorithms [6]. Kriby et al evaluated the accuracy of 11different DIR algorithms using mean spatial error and DSC [7].

A number of groups have been studied for evaluation of deformed dose using deformable physical phantom and various detectors. In case of two-dimensional detectors, Cherpak et al used MOSEFET for 4D dose-position verification in a deformable lung phantom [10] and Serban et al evaluated 4D radiotherapy verification using film [11]. Yeo et al used three dimensional dosimeters (GEL) to evaluate various algorithms [12].

In this study, we evaluated accuracy of various DIR algorithms using variations of the deformation point and volume in the virtual deformation QA software and various metrics.

## II. EXPERIMENTS

### 1. Virtual deformation QA software (ImSimQA software)

ImSimQA software includes virtual phantom library and manipulation of DICOM-3 images [13] and the Thin-Plate Splines (TPS) algorithm was used in ImSimQA software to carry out global deformation of volumetric image sets [14]. Varadhan et al studied about validation of DIR using this program [15] and Nie et al used virtual QA phantom provided in this program to evaluate accuracy of DIR algorithms [16]. It is possible to designate the fixed point and deformation point on the image and to move the deformation point from initial position [17]. We generated reference DICOM images from virtual phantom library and the DICOM reference images were deformed using global deformation of ImSimQA software. We acquired new DICOM images (deformed images)

### 2. Deformable image registration (DIR) algorithms

The open- source toolkit based MATLAB (Math Works, Natick) script was used for this study. The toolkit was called the Deformable Image Registration and Adaptive Radiotherapy Toolkit (DIRART) that classified into four DIR algorithms which the user can handle easily. The classes of the four DIR algorithms were following that Optical flow algorithms, Demons algorithms, Level-set algorithms and B-spine algorithm [18]. Four algorithms of two classes were selected for this study. One class was optical flow algorithms that were Horn- schunck (HS) [19] and Iterative Optical Flow (IOF) [20]. The other class was demons algorithms that were Modified Demon (MD) [21] and Fast Demon (FD) [22].

## 3. The evaluation of DIR algorithms for deformation point variations

The reference image (CT image of circle (d = 5 cm), 512×512, $I_{ref}$) was generated from the ImSimQA software. We deformed $I_{ref}$ using axial movement of deformation point and gained deformed images ($I_{def}$). As showed figure 1, a deformation point (red) was located in edge of the circle and the moving distance was from 3 mm to 30 mm (3, 5, 8, 10, 15, 20, 25 and 30). When deformation point was moved axially, $I_{ref}$ was deformed to $I_{def}$ by TPS algorithms of ImSimQA software [14].

The $I_{def}$ was inversely deformed to $I_{ref}$ by DIR algorithms and we acquired new image ($I_{id}$) from $I_{def}$. For evaluation of DIR algorithms, the image similarity between $I_{ref}$ and $I_{id}$ was calculated

## 4. The evaluation of DIR algorithms for volume variations

The reference volume (CT image, sphere (d = 5cm), slice thickness: 2.5 mm, 512×512×52, $V_{ref}$) was generated from the ImSimQA software. The $V_{def}$ was generated using deformation points located in a cube like figure 2. When deformation points were moved to lateral or medical direction, the $V_{ref}$ was deformed globally by TPS algorithms of ImSimQA.

The $V_{def}$ were classified into two depending on the type of deformation. The deformation 1 means that deformation points located in cube were moved to lateral direction (+) to relax the $V_{ref}$ and deformation 2 was that they were moved medial direction(-) to contract the $V_{ref}$. When the moving distance of deformation points was from -15 mm to 15 mm, the range of $V_{def}$ was from 28.64 cm$^3$ to 128.58 cm$^3$ (figure 3).

The $V_{def}$ was inversely deformed with respect to $V_{ref}$ by DIR algorithms and acquired the inversely deformed volume ($V_{id}$). The degree of overlap between $V_{ref}$ and $V_{def}$ and $V_{id}$ ratio to $V_{ref}$ were calculated to evaluate DIR algorithms.

## 5. The metrics for evaluation of DIR algorithms

(1) Normalized Mutual information (NMI)

The NMI was metric to measure image similarity- based image information [23].

$$\text{NMI}(A, B) = \frac{H(A) + H(B)}{H(A, B)}$$

Where H (A) and H (B) are the entropies of images A and B, respectively, and H (A, B) is their joint entropy. A: reference Image ($I_{ref}$), B: Inversely deformed Image ($I_{id}$)

(2) Normalized Cross Correlation (NCC)

The NCC was metric to measure image similarity-based intensity [24].

$$\text{NCC}(A, B) = \frac{\sum_{j=0}^{N-1} \sum_{i=0}^{M-1} (A(i,j) - \bar{A})(B(i,j) - \bar{B})}{\sqrt{\sum_{j=0}^{N-1} \sum_{i=0}^{M-1} (A(i,j) - \bar{A})^2 \sum_{j=0}^{N-1} \sum_{i=0}^{M-1} (B(i,j) - \bar{B})^2}}$$

Where A (i, j) and B (i, j) are the reference image ($I_{ref}$) and the inversely deformed image ($I_{id}$) of the coordinate (i, j), respectively. N and M represent the dimensions of the image matrix N×M.

$\bar{A}$ : The mean intensity value in reference image

$\bar{B}$ : The mean intensity value in inversely deformed image

2) Dice Similarity Coefficient (DSC)

To evaluate the similarity of volume, we calculated the degree of overlap between $V_{ref}$ and $V_{id}$ using Dice similarity coefficient (DSC) as the following: [24]

$$\text{DSC}(A, B) = \frac{2 \cdot |A \cap B|}{|A| + |B|}$$

A: reference volume ($V_{ref}$), B: Inversely deformed Volume ($V_{id}$)

## III. RESULTS AND DISCUSSION

**1. The evaluation of DIR algorithms for deformation point variations**

1) Image similarity ($I_{ref}$ vs $I_{id}$)

Figure 4 shows results of image similarity between $I_{ref}$ and $I_{id}$ depending on the moving distance of the deformation point in the all the DIR algorithms. The degree of image similarity was increased relatively after DIR applied to $I_{def}$. Since the moving distance of deformation point was increased, the degree of image similarity was decreased. When deformation point was moved to 4 mm, the value of NMI was above 1.81 and the value of NCC was above 0.99 in all DIR algorithms. The values were of them was maintained up to 8mm in DIR algorithms except for the IOF.

**2. The evaluation of DIR algorithms for volume variations**

1) The generation of $V_{id}$ by DIR algorithms

The $V_{def}$ was inversely deformed to $V_{ref}$ using DIR algorithms and acquired a new volume ($V_{id}$). Figure 5 shows the axial, sagital and coronal images of $V_{id}$ depending on the DIR algorithms and the optimized parameters of each DIR algorithms in DIRART were decided by the pre- study by Yeo et al [25].

2) The $V_{id}$ ratio to $V_{ref}$ and calculation of overlap between $V_{ref}$ and $V_{id}$

Figure 6 shows the $V_{id}$ ratio to $V_{ref}$. The closer the ratio is to 1, the performance of DIR algorithms was better. The performance of DIR algorithms was compared in deformation 1 depending on the moving distance of deformation points. The range of variation among algorithms was from 2.2% (moving distance of deformation points: 3mm) to 11.3% (moving distance of deformation points: 15mm). When the $V_{def}$ was increased 12% than $V_{ref}$ (moving distance of deformation points: 3mm), the difference between $V_{ref}$ and $V_{id}$ was an average of

4.4 %

In case of deformation 2, the variation range of among algorithms was from 3.4% (moving distance of deformation points: 3mm) to 17.9% (moving distance of deformation points: 15mm). When the $V_{def}$ was reduced 12% than $V_{ref}$ (moving distance of deformation points: 3mm), the difference between $V_{ref}$ and $V_{id}$ was an average of 4.3%

The value of DSC between $V_{ref}$ and $V_{id}$ was showed in figure 7. In deformation 1, the range of variation among algorithms was from 2.9% (moving distance of deformation points: 3mm) to 7.7% (moving distance of deformation points: 15mm). When the 12% of $V_{ref}$ was increased (moving distance of deformation points: 3mm), the value of DSC between $V_{ref}$ and $V_{id}$ was above 0.95 in DIR algorithms except MD algorithms. In case of deformation 2, the range of variation among algorithms was from 0.9% (moving distance of deformation points: 3mm) to 12.9% (moving distance: 15mm). When the 12% of $V_{ref}$ was increased (moving distance of deformation points: 3mm), the value of DSC between $V_{ref}$ and $V_{id}$ was above 0.95 in all DIR algorithms.

Kriby et al divided into three categories for evaluating the spatial accuracy of DIR. The methods are as follows: contour comparison, landmark tracking and simulated deformation. The landmark tracking methods were used widely [7]. The spatial accuracy of DIR was evaluated through the position change of landmark before and after applying the algorithm, For instance, Wognum et al calculated spatial error of DIR using the excise porcine bladders attached fiducial maker [6].

On the other hand, this study generated the $I_{ref}$ from virtual simple phantom of ImSimQA software and deformed $I_{ref}$ using deformation points of it. We easily could gain $I_{ref}$ and $I_{def}$ without image acquisition modality. The accuracy of DIR algorithms was estimated using the image similarity, the $V_{id}$ ratio to $V_{ref}$ and the degree of overlap between $V_{ref}$ and $V_{id}$

Yeo et al designed three types of deformation in order to evaluate the accuracy of DIR

algorithms [1, 25]. In this study, the reference volume was deformed into two types of deformation; relaxation and contraction. This pattern of deformation was designed in the light of volume variation of tumor and normal tissue during multi-fraction radiotherapy. For example, when a patient is treated for tumor in the pelvic region, the volume of bladder has potential to increase than volume in treatment planning.

Yeo et al certified the performance of DIRART algorithms and the results was best (HS) and worst (MD) [25], While Wognum et al calculated surface distance error (SDE) to evaluate accuracy of various DIR algorithms and the performance of IOF was worst in the all DIR algorithms[6]. As shown the result of previous study, the results of evaluation depend on the parameter of deformation and the deformation type.

The results of this study were best (HS) and worst (MD) in volume similarity with DSC. In case of NMI and NCC for evaluating image similarity, HS was best and IOP was worst.

The results according to NMI, NCC and DSC are different for each algorithm. That is because the value of NMI and NCC was calculated based on the image information and intensity, while the value DSC was calculated degree of overlap between the contours on the two images [23, 24].

The Trend of $V_{id}$ ratio to $V_{ref}$ was similar with it of DSC in almost all the algorithms. However in the MD algorithm, the value of DSC (0.89) was lower than it of another algorithm (0.95) in deformation1 (moving distance: 3mm) (Figure 7). The cause of this result was found in differences between sagital & coronal image of MD algorithms and them of reference (Figure 5).

As shown in our results, Though $V_{def}$ restored to $V_{ref}$ using DIR, it could be occurred discrepancy of volume similarity that would have a significant influence on the radiation treatment planning

## Ⅳ. CONCLUSIONS

Since the degree of deformation increased, the image similarity was decreased regardless of the type deformation. Four DIR algorithms were evaluated quantitatively and found out the following that when the $V_{ref}$ (63.28 cm$^3$) increased or decreased about 12%, the difference between $V_{ref}$ and $V_{id}$ was less than ±5%.

The DIR algorithms could not deform fully like $I_{ref}$ and $V_{ref}$ depending on the degree of deformation and the type of deformation. Hence, the performance of DIR algorithms should be verified for the desired application.

## ACKNOWLEDGEMENT

This work was supported by Ministry of Science, ICT and Future Planning and the author would like to thank D.K medical (Mr. Ju hyun Lee) for providing ImSimQA software.

Figure Captions.

Fig. 1. Deformed image ($I_{def}$) generated from ImSimQA software (The moving distance of deformation point (red) was located in edge of the circle was from 3 mm to 30 mm. and when deformation point was moved, image was deformed from reference image. (a) Reference image, (b) 8mm, (c) 30 mm)

Fig. 2. Deformed volume ($V_{def}$) depending on moving distance of deformation point that was located in the center of six surface in cubic and the $V_{def}$ was generated according to the type of deformation that were deformation1 (a) and deformation2 (b).

Fig. 3. Variation of $V_{def}$ generated by the global deformation of ImSimQA software depending on the type of deformation like figure 2 (The sign implied moving direction of deformation points and (+) sign was lateral, (-) sign was medial).

Fig. 4. An image similarity calculated between $I_{ref}$ and $I_{id}$ depending on the moving distance of deformation point in DIR algorithms (from 3mm to 30mm). The value of black rectangular was NMI and NCC before application of DIR algorithms (a) NMI, (b) NCC

Fig. 5. (a) Axial image, (b) Sagital image, (c) coronal image of $V_{id}$ in DIRART software, according to DIR algorithms (Red line: contour of $V_{ref}$, Blue line : contour of $V_{def}$, yellow line : contour of $V_{id}$), The $V_{def}$ was obtained when moving distance of deformation points was 3mm in defromation1.

Fig. 6. The $V_{id}$ ratio to $V_{ref}$ depending on the moving distance of deformation points in DIR algorithms. (The value of $V_{id}$ / $V_{ref}$ of various DIR algorithms means the $V_{id}$ ratio to $V_{ref}$ after application of DIR algorithms and the value of black rectangular was it before them)

Fig.7.The degree of overlap calculated between $V_{ref}$ and $V_{id}$ depending on the moving distance of deformation point in DIR algorithms. (The value of DSC means degree of overlap between $V_{ref}$ and $V_{id}$ after application of DIR algorithms and the value of black rectangular was it before them)

**Fig. 1.**

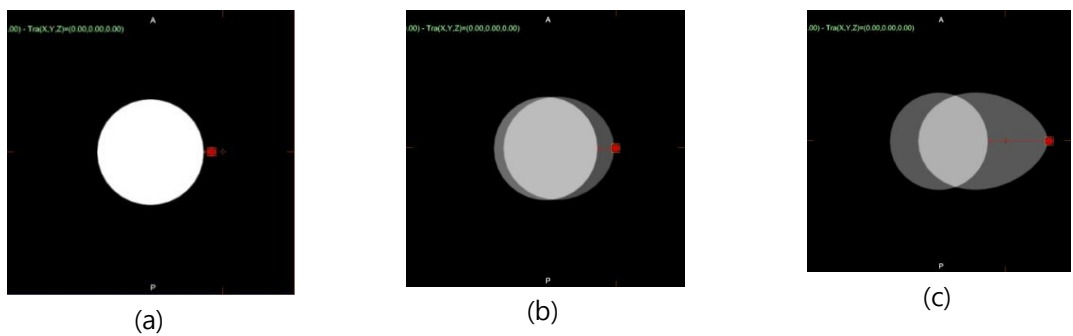

**Fig. 2.**

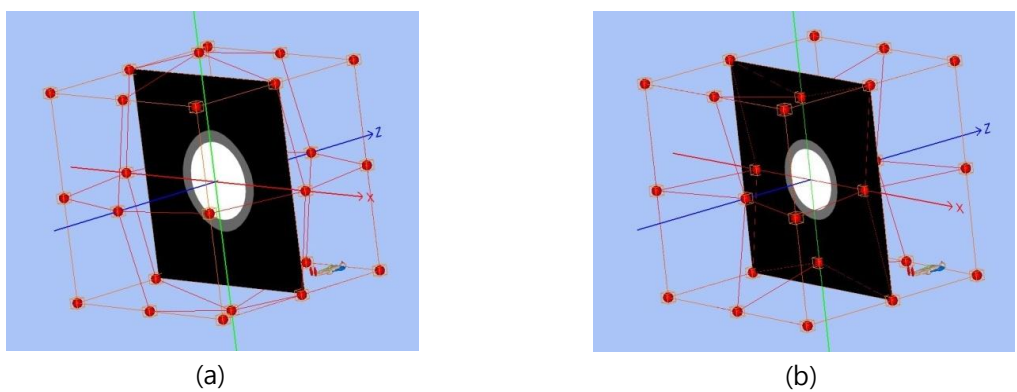

**Fig. 3.**

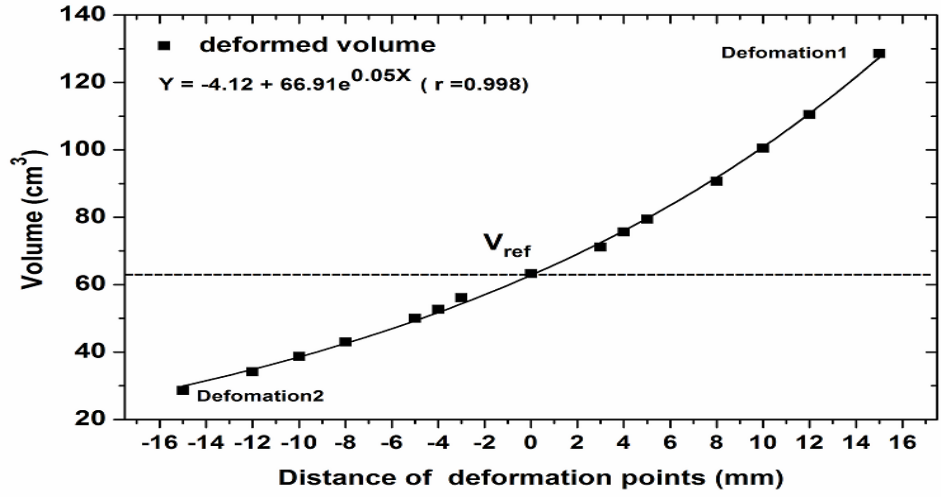

**Fig. 4.**

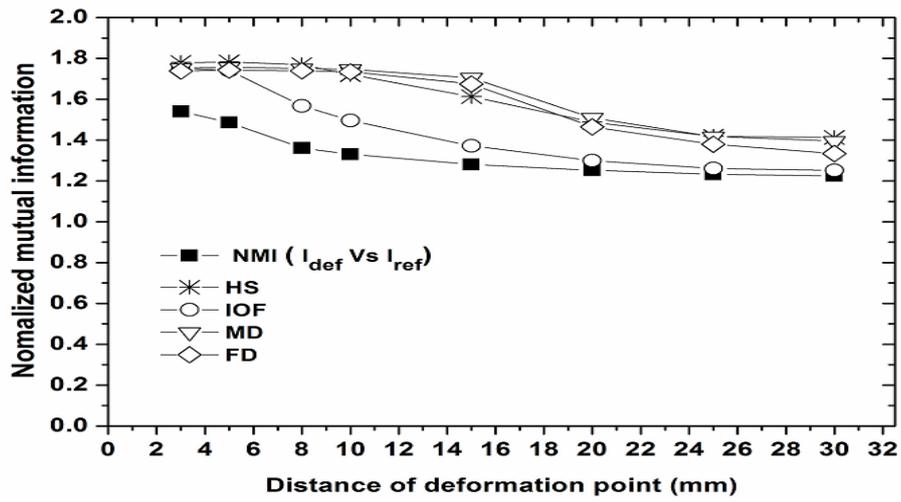

(a)

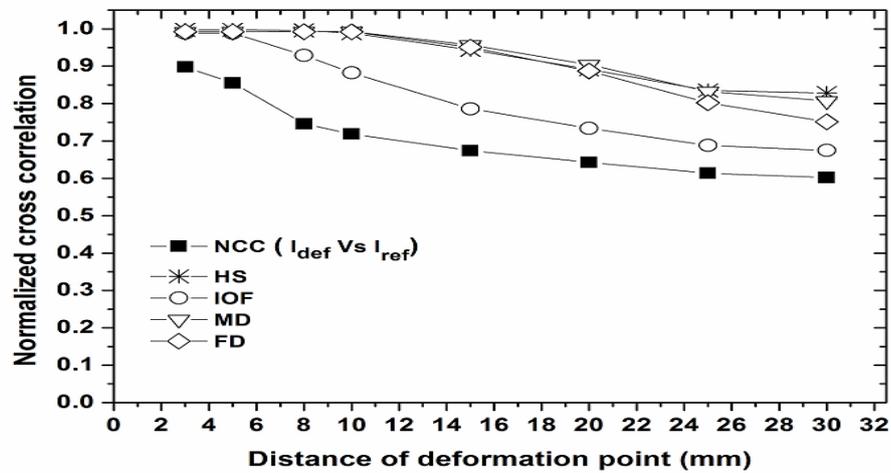

(b)

**Fig. 5.**

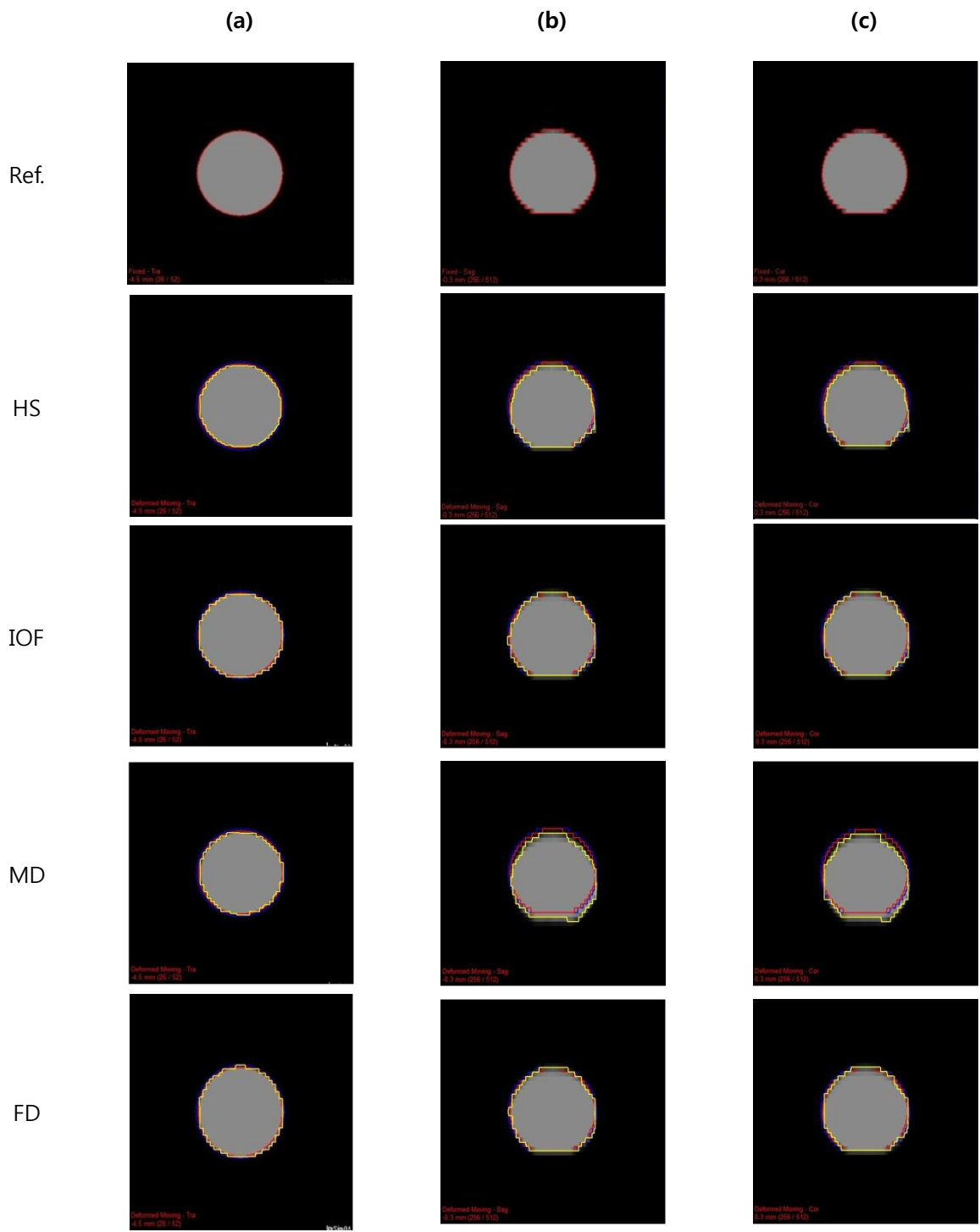

**Fig. 6.**

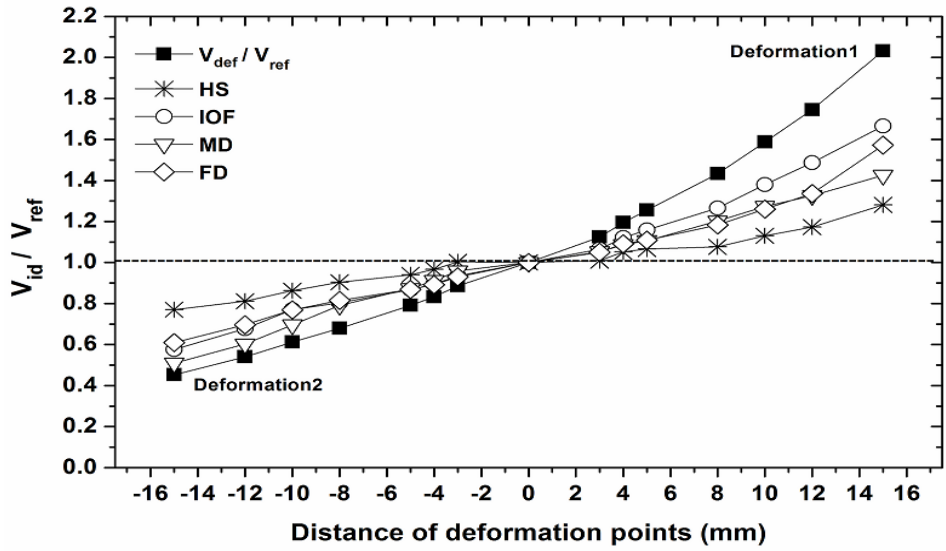

**Fig. 7.**

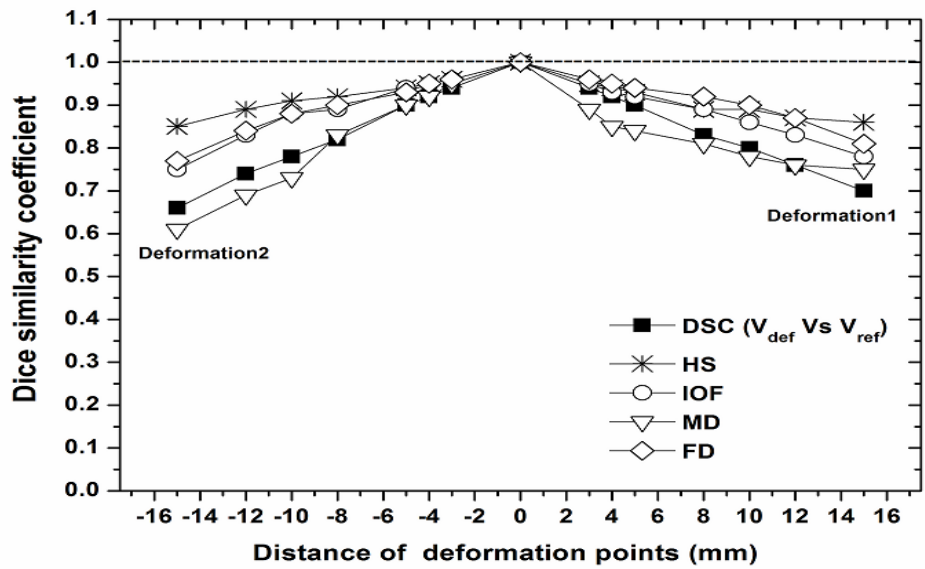